# Grid-Edge Energy-Flexible Technologies: A Comparative Analysis Across Generators, Loads, and Energy Storage Systems

Jesus Silva-Rodriguez[a], Tianxia Zhao[b], Ran Mo[c], Xingpeng Li[a]

[a]University of Houston, Department of Electrical and Computer Engineering, Houston, TX, 77204-4005
[b]Shell Energy North America, Houston, TX, 77002
[c]Shell International Exploration & Production Inc, Houston, TX, 77082

**Abstract**

This review analysis presents a comprehensive exploration of energy flexibility in modern power systems. It examines the roles and mechanisms of flexible technologies across three main categories: generators, energy storage systems (ESS), and loads. Energy flexibility is defined as the ability to dynamically adjust supply and/or demand in response to grid conditions to maintain balance and stability. This is of particular importance to facilitate the integration of the growing variable renewable energy sources (RES) into modern power grids. Additionally, traditional supply-side mechanisms to maintain balance and stability are complemented by advancements in demand-side management and demand response strategies, which enable loads to adjust consumption patterns and schedules in response to grid requirements. ESS are also explored to further enhance flexibility by absorbing excess generation and/or supplying large load increases that are not able to be met by the less flexible resources. This paper also explores specific flexibility technologies, examining their characteristics, control strategies, advantages, and limitations. Energy flexibility services are also categorized into intermittency mitigation, peak shaving, and energy reserve provisioning. Each service is supported by case studies and examples demonstrating how different resources respond to varying conditions. Ultimately, the findings and reviews of the various flexible resources in this paper provide a roadmap for optimizing energy flexibility across diverse resource types, paving the way for a more sustainable and resilient energy future.

*Keywords:* Demand Response, Energy Flexibility, Generation Control, Variable Generation, Renewable Integration.

## Abbreviations

| | | | |
|---|---|---|---|
| ADRC | Active Disturbance Rejection Controller | LHS | Latent Heat Storage |
| AGC | Automatic Generation Control | LSTM | Long-Short Term Memory |
| ANN | Artificial Neural Network | MDP | Markov Decision Process |
| ASTS | Active Solar Tracking System | MILP | Mixed-Integer Linear Programming |
| BAS | Building Automation System | MINLP | Mixed-Integer Nonlinear Programming |
| BDL | Bayesian Deep Learning | MISOCP | Mixed-Integer Second Order Cone Programming |
| BESS | Battery Energy Storage System | MOPSO | Multi-Objective Particle Swarm Optimization |
| CCGT | Combined-Cycle Gas Turbine | MPC | Model Predictive Control |
| CAES | Compressed Air Energy Storage | PAS | Process Automation System |
| CPS | Cylinder Pressure Sensor | PCM | Phase Change Material |
| CVS | Controllable Voltage Source | PCT | Programmable Communicating Thermostat |
| DER | Distributed Energy Storage | PHS | Pumped Hydro Storage |
| DP | Dynamic Programming | PLC | Programmable Logic Controller |
| DR | Demand Response | PSO | Particle Swarm Optimization |
| DSM | Demand-Side Management | RES | Renewable Energy Source |
| EDEALS | Electricity Demand-Response Easy Adjusted Load Shifting | RL | Reinforcement Learning |
| EIA | Energy Information Administration | RO | Robust Optimization |
| EMCS | Energy Management and Control System | RTN | Resource-Task Network |
| ESMS | Energy Storage Management Software | SHS | Sensible Heat Storage |
| ESS | Energy Storage System | SLURM | Simple Linux Utility for Resource Management |
| EV | Electric Vehicle | SO | Stochastic Optimization |
| FLC | Fuzzy Logic Control | STATCOM | Static Synchronous Compensator |
| GA | Genetic Algorithm | SVR | Supported Vector Regression |
| HES | Hydrogen Energy Storage | TCL | Thermostatically Controlled Load |
| HVAC | Heating, Ventilation, and Air Conditioning | V2G | Vehicle-to-Grid |
| ICE | Internal Combustion Engine | VLP | Variable Load Path |

# 1. Introduction

Energy flexibility refers to the ability to respond to changes in electricity demand and generation [1]. This is done to maintain a balance between supply and demand, which is required to keep system stability. Matching generation to demand in real time is referred to as energy balancing [2] and it is normally achieved by ensuring that sufficient generation capacity is always ready to be dispatched to meet any unforeseen demand changes.

Power balance has traditionally been achieved by only adjusting power generation on the supply side [3]. However, a diverse set of technologies on the demand side, such as automation control and smart devices, are now available, enabling demand-side resources to also respond to grid-changing conditions [4].

In addition, distributed energy resources (DER), particularly variable renewable energy sources (RES), can create balancing challenges for grid operators as well, since generation uncertainty is now included, and the grid was not originally designed for a diverse, decentralized mix of energy sources [2], [5]. Therefore, grid operators are now focusing on energy flexibility as a practical, cost-effective solution for balancing supply and demand [6].

Energy flexibility can be obtained from both generators and loads. Generators can implement power output adjustments in response to load variability, and flexible load demands can reduce consumption or shift its timing to match supply conditions [7]. Moreover, energy storage systems (ESS) can also be implemented to improve flexibility between generators and loads to either supply excess demand or absorb excess generation [8].

As a result of the increasing importance on energy flexibility to maintain grid stability, improve system reliability, enable seamless integration of RES, and improve overall power system efficiency, this paper provides a comprehensive review of the different sources of energy flexibility across generators, loads and energy storage systems, on both supply and demand sides of the grid. The different characteristics, control strategies, and technologies that make their flexible operations possible are particularly highlighted, in addition to collecting a variety of generators, loads, and storage devices as examples to investigate how well they may perform in responding to various supply/demand changing conditions.

*1.1. Literature Review*

Various papers and articles in the literature have explored energy flexibility on both supply and demand sides of the grid, in particular to accommodate the ever-increasing share of variable RES. In [6], increases in uncertainty and variability due to the inclusion of large variable RES have been identified, and the need for efficient computational algorithms for future power system planning incorporating energy flexibility has been especially highlighted. Moreover, the concept of energy flexibility, considering measures and innovations on both the supply and demand side, has been presented in [3], encompassing multiple resources such as controllable generators like coal and gas plants, DER, and flexible demand and energy storage.

Quantification and mathematical modeling of energy flexibility have been explored as well, such as in [9], where a mathematical formulation is proposed to quantify flexibility provisions and characteristics such as power operating range, instantaneous power, energy storage capacity, and response duration. Moreover, [10] and [11] explore different energy flexibility sources on the demand side, particularly in grid-responsive buildings, where flexibility is quantified as capacities and ratios that reflect how much contribution a building can make based on flexible load characteristics such as regulation, load shedding, load shifting, and load covering. Additionally, levelized cost quantification of energy flexibility is considered in [12], where the total cost per unit of energy flexibility capacity during the lifetime of the resource is factored in, and used as comparison with other demand-side flexible technologies using metrics based on different costs such as investment, operation, and replacement costs.

As previously stated, there has been a substantial increase in interest in technologies that enable demand-side flexibility. These technologies may range from direct energy consumption management of net loads and thermal loads, but also from onsite distributed generation and energy storage that may be present [11]. A model that integrates building-integrated and vehicle-integrated photovoltaics, as well as battery energy storage systems (BESS) and electric vehicle-to-grid (V2G) responses, is presented in [13], which is used to optimize flexibility responses to grid-changing conditions such as peak shaving and load shifting, and [14] explores the use of phase change materials (PCM) inside the envelope of a building to increase thermal mass that can be used to increase load shifting capabilities of HVAC systems. Both of these papers explore ways in which load resources can be used to offer flexibility and facilitate the integration of RES.

In addition to loads, some generators can offer flexibility as well, and flexibility can be even further enhanced with the integration of energy storage technologies that supplement flexibility for both loads and generators. A study characterizing low-carbon power generation and long-duration energy storage technologies is presented in [15], which shows that many combustion-based generators such as combined-cycle plants have experienced improved fast-start and load-following capabilities, but concludes that energy storage is the most viable form of flexibility to complement variable RES, such as hydrogen storage and compressed air. Similarly, [8] provides an overview of the reliability impacts of ESS combined with other cost-effective flexible technologies such as demand response (DR), which shows how ESS can improve continuity of supply and achieve an optimal utilization of grid resources via time shifting, output smoothing, and transmission congestion mitigation, among other benefits.

While these reviews and articles identify different sources of flexibility on both supply and demand side as well as what characteristics a resource needs to be flexible, they offer limited examples of specific flexible resources available, and most studies are performed from a demand-side perspective, considering mostly DER, DR and demand-side management (DSM) technologies. Moreover, there is a lack of direct comparison among supply and demand side resources, and in particular, among generators, loads, and energy storage systems, that can serve to identify the most flexible technologies or even the best combination of resources to offer the most flexibility to the grid. Therefore, this paper explores energy flexibility capabilities for different types of generators, loads, and ESS in order to explore the current possibilities that exist to provide flexibility to the power grid and be able to accommodate the continuous integration of variable load and generation to the system.

*1.2. Structure of the Review*

The remainder of this paper is structured as follows. First, Section 2 presents a categorization of different energy-flexible resources, provides background and fundamental information on each resource type, and identifies different key characteristics of each resource to ultimately draw a comparison among all resources considered within their respective categories. Section 3 then presents a characterization of energy flexibility by selecting different examples of energy-flexible resources from each category and analyzing their performance when providing flexibility services such as intermittency mitigation, peak shaving, and energy reserve, based on their flexible parameters such as power capacity, ramping rate, and response time and duration, obtained from extensive literature research. Lastly, Section 4 concludes the paper.

## 2. Categorization of Energy-Flexible Resources

In the context of this paper, an energy resource is defined as any component that directly contributes to the general supply and/or demand of electricity in the network. Thus, it is not exclusive to sources that generate electricity, but also the consumers of electricity. On the supply side, there are generators; and on the demand side, there are loads; and energy storage can act as both [5]. Thus, energy storage can be a part of energy suppliers or consumers.

Energy flexibility capabilities for all three of these resource types are explored in this section for different generators, loads, and energy storage systems. This exploration is performed to identify existing current possibilities to provide energy flexibility to the power grid and be able to accommodate the continuous integration of variable load and generation to the system with little interruption. Table 1 shows the different resources that are reviewed and analyzed, categorized by resource type.

Table 1: Energy flexibility resources, analyzed and quantified in this paper, by type and category.

| Resource Category | Resource Type | Flexible Resource Asset |
|---|---|---|
| Generation | Thermal Generation | Internal Combustion Engine (ICE) |
| | | Combined-Cycle Gas Turbine (CCGT) |
| | Renewable Generation | Hydropower Plant |
| | | Solar PV |
| | | Wind Turbine |
| Load | Thermostatically Controlled Load (TCL) | Heating, Ventilation, and Air Conditioning (HVAC) |
| | | Refrigeration |
| | Industrial Processes | Cement Production |
| | | Metalworking |
| | | Oil Refinement |
| | Computing Resources | Data Centers |
| Energy Storage | Electrochemical Energy Storage | Batteries |
| | Chemical Energy Storage | Hydrogen Storage |
| | Mechanical Energy Storage | Flywheel |
| | | Pumped Hydro |
| | | Compressed Air |
| | Thermal Energy Storage | Latent Heat |
| | | Sensible Heat |

When reviewing the energy flexibility of these resources, this paper focuses on the following characteristics to understand how they are able to provide energy flexibility:
- **Flexibility Service:** Type of energy flexibility action offered by the flexible resource.
- **Flexibility Action:** Actual response taken to provide energy flexibility.
- **Optimization Strategy:** Computing tool or algorithm type used to implement the flexibility action.
- **Supplement Technology:** Computing hardware and/or software solutions used to implement the control system/strategy.
- **Advantages:** Benefits obtained by the implementation of this flexibility action.
- **Disadvantages:** Drawbacks encountered by the implementation of this flexibility action.
- **Response Time:** Typical time required for the flexibility resource to fully implement its flexibility action.
- **Response Duration:** Typical duration time of the flexibility action implemented.

*2.1. Generators*

Flexibility actions that can be taken by generators are limited to ramp-up/ramp-down actions, as well as start-up/shut-down actions [6]. However, flexibility services for generators could be defined as the types of ancillary services. According to the Energy Information Administration (EIA), ancillary services ensure reliability and support the transmission of electricity from generation sites to customer loads, and include such services as load regulation, spinning reserve, non-spinning reserve, replacement reserve, and voltage support [16]. Therefore, ancillary services could be categorized under energy shifting, energy reduction, and so on.

Flexibility services for generators that encompass these ancillary services could then be defined as the following:
- **Output Regulation:** Quick response to instantaneous changes in electrical demand by increasing or reducing power output within a couple of seconds.
- **Online Generation Reserve:** Online generator maintained at an output rate that allows for a substantial and fast increase in generation as a backup, ideally more than half its capacity within a matter of minutes.
- **Offline (Supplemental) Generation Reserve:** Generator maintained in an offline state that can start up and ramp up within a matter of minutes to provide backup power when needed.

The flexibility actions and all other characteristics of each generator analyzed in this paper are presented based on how they can provide these services. Additionally, Table 2 summarizes all the information pertaining to the generators considered in this section, and how they can provide these three main flexibility services.

*2.1.1. Thermal Generators*
Thermal generators are fully controllable generators that depend on a consumable fuel, such as coal or natural gas, that is ignited to convert the released thermal energy into electrical energy. Some of the most flexible thermal generation technologies are the combined-cycle gas turbine (CCGT) and the internal combustion engine (ICE).

*2.1.1.1. Combined-Cycle Gas Turbine*
CCGTs are well suited for flexible power generation, with possibilities for steep power output gradients and quick start-up of gas turbines. Moreover, with the ability to operate in various modes, the bottoming steam cycle presents additional opportunities for a wide and flexible load range. Its operational flexibility is given by its minimum power output level, power ramp rate, and cycling characteristics, while its fuel input flexibility is given by regulating steam flow and hence varying the ratio of electricity output [17].

Additionally, supplementary firing is a technology that can improve peak power and load flexibility, thus serving as a way of increasing net plant power output by installing duct burners in the heat recovery steam generator. However, there are temperature restrictions that limit its prolonged use [18].

Common control strategies to implement the most efficient operation revolve around optimization methods for unit commitment and effective control of the different possible flexibility actions. Some examples of these optimization models are mixed integer linear programs (MILP) that facilitate unit commitment, mixed-integer second-order cone programming (MISOCP) for more efficient iterative methods for operation scheduling, and dynamic programming (DP) for better power output dispatch and regulation [17], [19]. In addition, machine learning methods such as support vector regression (SVR) have shown appropriate prediction capability for determining optimal maximum power operating points as well as efficient performance [20].

*2.1.1.2. Internal Combustion Engine*
Although mostly known to be implemented for the transportation industry, ICE technology has been rapidly developing into a plausible option in large-scale power generation, displaying notable synergies with renewable energy generators. Renewables and ICEs can unlock greater cost savings together than either technology on its own [21]. ICEs can provide fast-reacting operations that can ramp up from zero to 100% output in less than five minutes, regardless of plant size. Moreover, ICEs can increase fuel security through the capability to burn any gaseous and liquid fuels to generate electricity, in addition to having short construction times that are typically less than one year [21]. However, air quality and associated emissions must be managed in accordance with local regulations.

ICEs in modern power systems are capable of managing air-fuel ratios through advanced control systems and the use of technologies such as turbochargers that allow air pressurization to maintain the correct air-fuel ratio. Moreover, an ICE system can implement sequential start-up and shut-down actions quickly via the deployment of individual units, allowing for flexible power output. In this way, each engine operates independently, making it possible to turn on or off individual units at any time without affecting the output of other units, and also achieve synchronized power ramping [22], [23]. Additionally, the use of multiple engine units combined with each unit's modular design allows for efficient operation of the entire system even at low power output levels [23].

Optimization for flexible operation of ICEs can be effectively achieved by MILP and DP models that allow optimal scheduling, dispatch, and power output control of each engine unit [24]. Additionally, machine learning models such as SVR and artificial neural networks (ANN) have been used for extensive modeling of ICE control and performance optimization [25].

*2.1.2. Renewable Generators*

Although not very well known for being substantial sources of energy flexibility, renewable generators can still offer flexibility via automatic control of their operating points. This is particularly true of solar and wind power generators that, while their maximum power output mostly depends on the available resource, deliberate manipulation of the maximum power point tracking (MPPT), for example, can allow the generator to operate at a lower point which allows it to offer an increase later on if required [26]. However, in the case of hydropower, this is a very mature and proven generation technology that is controllable and very flexible.

*2.1.2.1. Hydropower Plant*

Hydropower is more extensively utilized for hourly ramping flexibility than any other resource. The ability to change generation in this system depends largely on the release decisions of its individual reservoirs, and there are numerous constraints imposed by the general reservoir operating rules that affect these decisions [27].

Hydropower plants control water flow through reservoir management, turbine gates, and the use of spillways. By adjusting turbine release and spill rates, hydropower plants can optimize their power generation while meeting environmental constraints. This process provides the plant with the ability to provide energy reserve services. Also, hydropower units can start and stop very quickly, allowing the plant to rapidly respond to variable power demands. However, frequent start-stop operations can increase wear and tear on components like turbines, gates, and valves, thus limiting the frequency with which these actions can be implemented [27], [28]. Additionally, hydropower turbines can also be operated in a speed-no-load condition, where the turbine runs without generating electricity, but remains synchronized with the grid. This mode allows the plant to be prepared for quick activation when additional power is needed. However, while operating in speed-no-load helps minimize delays in starting up power generation, it can also result in increased cavitation and mechanical stress, particularly at lower gate openings, limiting the time the plant can operate in this state [28].

Flexible operation of hydropower plants performing these flexibility actions can be achieved using optimization strategies such as MILP as well as mixed-integer non-linear programming (MILNP) methods to maintain optimal operation when performing output regulation as well as acting as energy reserve. Moreover, ANNs can also be used to determine the best power operating point in different situations, depending on demand, and methods such as genetic algorithms (GA) can enhance water flow rate regulation responses to quickly follow grid-changing conditions [27], [29].

*2.1.2.2. Solar PV*

Solar PV systems generate power depending on solar irradiance available at any given time of the day. This makes it a rather variable generator, with power output that is not very flexible due to this limitation. Additionally, since solar PVs do not require any fuel, and the maximum power output possible is desired at all times to take advantage of the essentially free solar resource, then their power output in most cases is assumed to be uncontrollable. Nonetheless, if energy flexibility is necessary for power system stability, PVs can be made to operate at below nameplate power output levels, reducing generation lower than their current maximum power available [26].

Active solar tracking is used in PV systems to maximize the amount of solar irradiance absorbed by the panels at any time of the day, depending on the solar position in the sky. Therefore, the solar tracking system can be controlled to offset this optimal tracking and instead position the PV panels in a way that maximum power is not obtained, effectively reducing power output if required, or purposefully operating at a lower point to provide energy reserve [30]. Active solar tracking systems can be used along with multiple control algorithms, such as neuro-fuzzy logic or robust optimization (RO), as well as machine learning techniques such as ANN or Bayesian deep learning (BDL) [30].

Table 2: Energy-Flexible Generators [17]-[34].

| Flexibility Service | Resource Type | Flexibility Asset | Flexibility Action | Optimization Strategy | Supplement Technologies | Advantages | Disadvantages | Response Time | Response Duration |
|---|---|---|---|---|---|---|---|---|---|
| **Output Regulation** (Quick adjustment in power output) | Thermal Generators | CCGT | Steam flow regulation | DP, MILP, SVR | AGC | Various amplitudes of instantaneous flexibility | Waste of excess steam, Increased cost of preparing water for increased steam | 1 – 60 min | < 4 hrs |
| | | ICE | Air-fuel ratio control | DP, MINLP, ANN, SVR | Turbochargers, CPS, Throttle valve | Fast operation without compromise on efficiency | Acoustic noise pollution, air quality-related emissions | < 1 min | NO LIMIT |
| | Renewable Generators | Hydropower | Water flow rate regulation | ANN, MINLP, GA | Digital Turbine Governor, Level transmitter, Opening time limiter | Supports fast response to grid fluctuations due to variable sources | Deviations from optimal efficiency | < 1 min | NO LIMIT |
| | | Solar Power | Solar tracking operation | BDL, ANN, Neuro-fuzzy logic, RO | ASTS, AGC | Slight reduction in generation invariability | System efficiency reduction | < 1 min | 1 – 15 min |
| | | Wind Power | Blade pitch control | ANN, FLC, RO | Electric pitch controller, Soft computing controller | Efficient power control, Slight reduction in generation invariability | Extra complexities and increased costs | < 1 min | 1 – 15 min |
| **Online Generation Reserve** (Additional backup power from an already-operational unit) | Thermal Generators | CCGT | Supplementary firing burners | DP, MISOCP, SVR | VLP | Enables plant to remain on standby even when shutdown, Peak load increases | Efficiency reduction due to high fuel-gas temperatures, High-pressure steam could exceed limits | 30 – 120 min | 2 – 4 hrs |
| | | ICE | Low output continuous idle operation | MILP, DP | AGC, PLC | Almost immediate response to demand increases | Prolonged inefficient operation | < 1 min | NO LIMIT |
| | Renewable Generators | Hydropower | Speed-no-load operations | MILP, MINLP, ANN | Level transmitter, Opening time limiter | Large amount of potential supplemental power | Reduced fatigue life of turbines | 1 – 3 min | NO LIMIT |
| | | Solar Power | Curtailment | SO, RO, MILP, SVR | AGC | Prevent over-generation, Offer grid-stability services | Waste of renewable resource availability | 1 – 30 min | NO LIMIT |
| | | Wind Power | Curtailment | SO, RO, MILP, SVR | AGC | Prevent over-generation, Offer grid-stability services | Waste of renewable resource availability | 1 – 30 min | NO LIMIT |
| **Offline Generation Reserve** (Supplemental backup capacity for immediate dispatch) | Thermal Generators | ICE | Sequential start-up and shut-down | MILP, ANN | AGC, PLC | Fast and reliable backup response | Increased risk of voltage and frequency fluctuations related issues | 30 sec – 5 min | NO LIMIT |
| | Renewable Generators | Hydropower | Turbine quick start/stop operation | MILP, MINLP, ANN | Gate/Valve automatic control | Large amount of potential supplemental power, Fast and reliable backup response | Increased vibrations and dynamic stresses | 1 – 3 min | NO LIMIT |

Curtailment is another strategy that can be implemented to reduce solar power output when needed. Curtailing excess solar power prevents grid overloads and minimizes the need to ramp down other more inflexible generators [31], [32]. Solar plants may be equipped with control units that enable transmission system operators to remotely reduce the solar plant's power output via automatic generation control (AGC) systems. This achieves fast, real-time adjustments without requiring significant downtime [32]. Different control strategies and algorithms can be used to carry out solar power curtailment successfully. Among these are stochastic optimization (SO), RO, MILP, and SVR, which allow the implementation of optimal dispatch models compatible with variables RES [31].

*2.1.2.3. Wind Turbine*

Similar to solar PV systems, wind turbines heavily rely on the available wind speed, which varies stochastically over time. Generating as much power as possible with the available wind resource is also desired in this case, as this can also be considered a technology that uses a rather free resource. Therefore, while operating at a suboptimal point may not be the most ideal case, if required for flexibility purposes, the power output can be manipulated and deliberately reduced to comply with the grid stability requirements if necessary [33].

Blade pitch control for wind turbines is used to achieve smooth power generation by controlling the input torque of the incoming wind. Using an active control system along with the pitch angle control can vary the turbine's pitch angle to decrease torque and rotational speed, effectively reducing power output [34]. Different control strategies that can be used to implement blade pitch angle control involve optimization techniques such as ANN and SVR. These control strategies focus on controlling and regulating parameters such as rotor speed, output torque, and terminal voltage to appropriately determine the required power output as well as maximum power available from the current wind speed, and control pitch angle accordingly [34].

Power curtailment can also be implemented in cases where large amounts of power generation must be reduced in response to grid supply/demand imbalances. Similar to solar power curtailment, wind power curtailment can also be achieved with AGC systems to allow grid operators to control their output in real-time when they determine that there is an excess in electricity generation. Additionally, curtailment can also be implemented deliberately to operate in a dynamic reserve capacity and increase power output only when needed, thus ensuring operational reserves remain available without destabilizing the grid [31], [32].

*2.2. Loads*

Flexibility can also be provided by loads. This is done by varying their power demand to help match the current supply. The different flexibility actions of loads are possible thanks to DSM and DR strategies [6]. These enable consumers to communicate directly with the grid and modify their electricity consumption accordingly [2].

Flexibility services for loads are then possible due to different DR actions, which can be cataloged as follows:
- **Load Shifting:** Adjustment of electricity consumption by shifting demand from peak hours to off-peak hours.
- **Reversed Flow:** Bidirectional power exchange in which a load can store energy that can be later returned to the grid.
- **Load Reduction:** Temporary adjustment of electricity consumption to a lower operating point.
- **Load Removal:** Complete shutdown of a load to decrease excess demand.

The flexibility actions and all other characteristics for each load analyzed in this paper are presented based on how they can provide these services. Additionally, Table 3 summarizes all the information pertaining to the loads that are considered in this section, and how they are operated to provide energy flexibility via load shifting, load reduction, load removal, and, in the specific case of electric vehicles (EV), reversed flow services.

*2.2.1. Thermostatically Controlled Loads*

Thermostatically controlled loads (TCL) are those whose main purpose is to convert electricity into heating or cooling energy, such as Heating, Ventilation, and Air Conditioning (HVAC) systems and refrigerated warehouses. These loads are controlled by thermostats that set a desired temperature setpoint and temperature thresholds, and measure the interior temperature for control purposes [35]. Leveraging the thermostat technology as well as different control strategies and technologies, these loads can be capable of providing energy flexibility services.

#### 2.2.1.1. Heating, Ventilation, and Air Conditioning

HVAC systems have proven to be an excellent DR resource. First, they make up a substantial portion of the electrical load in buildings. Second, their thermal storage effect of indoor environments allows HVAC loads to be temporarily shed without immediate impact on building occupants. Third, HVAC systems are commonly at least partially automated with energy management control systems (EMCS) [12], [36], [37]. Moreover, another supplemental technology that allows HVAC systems to offer energy flexibility is their programmable communicating thermostats (PCT), which allow for effective ways of controlling temperature setpoints and exploiting the thermal mass in their interior to reduce or shift power consumption [36], [38].

Control strategies for load control of HVAC systems range from optimization models used to control scheduling as well as real-time responses, such as MILP and MINLP methods, DP, and GA [38]-[40], and machine learning techniques such as SVR, ANN, and reinforcement learning (RL), which are also used to aid in scheduling and control tasks while taking into account user preferences, such as preferred temperature set points and temperature thresholds [40], [41].

#### 2.2.1.2. Refrigeration

Cooling and refrigeration offer energy flexibility primarily through their excellent ability to exploit the inherent thermal inertia of the interior of refrigerated rooms and warehouses. Since these cooling processes are not too sensitive to short-term load curtailment, then DR activities do not affect them much [36], [42].

The operation and overall objectives of refrigeration loads are very similar to those of the HVAC. As a TCL itself, its objective is to maintain a certain temperature range, with these setpoints directly determining the amount of energy flexibility the refrigeration load will be able to provide [40], [42], [43]. As such, many of the same control strategies and supplemental technologies of the HVAC are also used.

### 2.2.2. Industrial Processes

The industrial sector is responsible for consuming more than half of the world's total delivered energy [44]. Therefore, DR and DSM on industrial processes have the potential to deliver the most energy flexibility for grid operators. However, operation scheduling of these processes must be done considering the complicated processes each industrial load type conducts, along with their hard constraints [44].

There are many industrial processes that are energy-intensive and can benefit from DR and DSM, among which some of the most dominant industries are metalworking, cement production, and oil refineries [42]. Each of these and the methods implemented to provide flexibility are presented in this subsection.

#### 2.2.2.1. Metalworking

Metalworking mainly involves processes for smelting and refining minerals to produce different pure metal or metal alloy products. The two most heavy metal industries are aluminum and steel [44].

Aluminum smelting pots operate at very high temperatures to maintain the aluminum that is produced in a molten metal form. However, the power consumption of a pot line can be adjusted very quickly by changing the output voltage of the rectifier that supplies a DC current to the pot line. Each pot has a large amount of thermal mass with a multi-hour thermal time constant that allows for instantaneous electric power variations that do not greatly impact the pot's thermal balance [42]. Moreover, melting scrap steel is also a large energy-consuming process in which heat is generated by electric arc furnaces or by means of induction, which lets the scrap metal start melting in the furnace. Although the process is able to halt instantly, the melting process must be resumed again before the disruption exceeds half an hour, since the scrap metal begins to cool down after that [42]. Therefore, this process can offer energy flexibility by completely removing the load, but only for small periods of time.

The optimization strategies for metal industries such as aluminum and steel often focus on the stochasticity of grid and electricity market conditions to predetermine how and when their flexibility actions may be required beforehand [42]. Some of these optimization strategies involve MILP models and resource-task network (RTN) models [42], [45], [46]. Machine learning techniques such as ANN and RL can also be used to better deal with the scheduling and control tasks of smelting and furnace units, as well as to better deal with the stochastic elements of the optimization process [41].

#### 2.2.2.2. Cement Production

The cement industry is one of the major consumers of energy, consuming annually over 350 trillion BTU of fuel and 10 billion kWh of electricity in the US [42]. A massive potential of DR in cement plants is found in huge grinding mills, especially in non-continuous processes like quarrying operations, raw mills, clinker mills, and fuel mills. Moreover, the ability to perform load shifting in these plants is mainly influenced by the automation level, scheduling, and storage abilities at each step [42].

The production process in cement plants can be modeled to minimize electricity costs with an optimization model, particularly particle swarm optimization (PSO), which can achieve optimal scheduling by mostly modifying the grinding operations. However, crushing and homogenization processes can also be shifted [44].

In addition to PSO, MILP models are also used, along with a dual robust variant of the MILP model denominated as R-MILP, and leveraging model predictive control (MPC) frameworks from these units [44], [47]. ANN models can also be used to better estimate modified and rescheduled grinding operations of the cement vertical grinding mill according to current grid conditions [48].

*2.2.2.3. Oil Refinement*

Oil refineries are complex processing facilities for transforming crude oil into marketable refined products. These are large facilities that involve substantial energy-intensive processes [49], making this an industry with a large energy consumption, with the potential for load management that can primarily provide load-shifting services.

The main flexibility action these load types can take is in the form of mass flow regulation, which allows adjustment of the crude oil input to the refinery unit, effectively changing power consumption, but also affecting production [49]. These actions are achieved via EMCS or MPC technologies, which implement optimization models such as MILP and PSO to determine production scheduling and material inflow regulation according to current grid conditions and production requirements [49]-[52]. Moreover, machine learning techniques such as SVR, ANN, and adaptive neuro-fuzzy inference (ANFI) models are also used to boost accuracy and optimize both production and energy consumption, depending on current flexibility requirements from the grid [52].

*2.2.3. IT Industry*

While IT was mostly embraced by industrial and manufacturing processes in the past, recent advances in networking, data storage, and AI have made this category establish itself today as an individual industry on its own, which has its own unique impact on the electricity demand [44]. Particularly, data centers are the IT loads that have grown the most in the past few years, and are projected to continue to grow substantially [53].

Nowadays, multiple data centers are industrial-scale operations using huge amounts of energy [44]. The IT sector is also projected to contribute about 50% to the total consumed electrical power, approaching 3,200 TWh over the next few years [54].

Workload shifting can be achieved in data centers utilizing computing resource management tools such as the Simple Linux Utility for Resource Management (SLURM) workload manager to perform job scheduling and manage job queues, reducing peak power demand and flattening the data center's demand curve [44], [55].

The SLURM workload manager, which allows for flexible computing job scheduling and power capping by halting and postponing computing tasks, is programmed via electricity demand-response easy adjusted load shifting (EDEALS) algorithms, as well as real-time optimization methods such as DP [55]. Moreover, ANN can also be used to enhance computing job scheduling that more accurately follows the flexible needs requested by the grid. RL can be used as well, leveraging the data center's capability for data-driven approaches that can implement demand response actions more seamlessly [41], [55].

*2.2.4. Transportation*

The introduction of EVs has added a new electrical load demand to the system that was not present previously. While these loads do not currently represent a huge share of the world's electricity consumption, these loads can be considered some of the most flexible loads in the grid [12], [56], in addition to EVs and EV fleets being expected to be widely adopted in the future [57].

EVs can provide energy flexibility to the grid by shifting their charging loads to less congested times, just like other load types such as TCLs. However, EVs are also unique in the sense that they can also discharge power from their batteries when needed, where this functionality is enabled in the vehicle and the local grid. This action is known as V2G response [13]. Thus, EVs can partially serve as distributed energy resources that can both consume and supply energy. This capability can help balance the grid by reducing demand during peak hours and providing energy during shortages [56], [58].

Various optimization models can be used to manage EV participation in demand response. The key optimization is to minimize the cost of energy while flattening the load curve. This is typically achieved by considering the EV charging profiles, battery state of charge, and user travel requirements. These are achieved by optimization methods such as MINLP models and DP models that allow scheduling and real-time response [56], [58].

Machine learning algorithms can also play a significant role in predicting the correct performance of EV charging, meeting both the flexible needs of the grid as well as user requirements. These methods allow for a more dynamic resource allocation optimization for more accurate planning and real-time adjustment to the charging and discharging activities. These can effectively be achieved with SVR and RL models for the correct charging scheduling and required charge levels at all times [41], [58].

Table 3: Energy-flexible loads [36]-[52], [54]-[56], [58]-[59].

| Flexibility Service | Resource Type | Flexibility Asset | Flexibility Action | Optimization Strategy | Supplement Technologies | Advantages | Disadvantages | Response Time | Response Duration |
|---|---|---|---|---|---|---|---|---|---|
| **Load Shifting** (Requires compensation in previous or subsequent hours) | TCL | HVAC | Passive thermal mass storage | DP, GA, SVR, RL, ANN | EMCS, PCT, BAS | Substantial potential for flexibility, Ease of implementation | Risk of rebound peak, Potential temperature discomfort of some users | 1 – 5 min | 1 – 4 hrs |
| | | Refrigeration | Passive thermal mass storage | DP, SVR, RL, ANN | EMCS, PCT | Substantial potential for flexibility, Ease of implementation | Risk of rebound peak, Risk of impact on product quality | 10 – 20 min | 30 min – 2 hrs |
| | Industrial Processes | Metalworking | Smelting heat rate adjustment | RTN, ANN, RL | EMCS, CVS | Fast response to DR signals | Response duration is short, Risk of unwanted prolonged interruptions | 1 – 5 sec | 30 min – 1 hr |
| | | Cement | Process rescheduling, Modified griding operations | PSO, MILP, ANN | EMCS, PAS | Significant potential cost reductions | Slow response (requires pre-planning), Potential loss of production | 5 – 10 min | 1 – 4 hrs |
| | | Oil Refinement | Process rescheduling | MILP, SVR, ANN, ANFI, PSO, GA | EMCS, MPC | Significant potential cost reductions | Process disruption | 2 – 4 hrs | 2 – 12 hrs |
| | IT Industry | Data Centers | Flexible computing job scheduling | EDEALS, ANN | SLURM workload manager | Fast response to DR signals, Ease of implementation, Potential long duration response | Potential decrease in resource use efficiency | 1 – 10 min | 30 min – 2 hrs |
| | Transportation | EV | Smart Charging | MILNP DP, SVR, RL | EMCS, CVS | Energy cost savings, Substantial potential for flexibility | Risk of rebound peak, User behavior, and preferences | 1 – 10 sec | 1 – 6 hrs |
| **Reversed Flow** (Load with bidirectional power flow capabilities) | Transportation | EV | V2G response | MILNP, DP, SVR, RL | EMCS | Grid support and stability, Fast response to DR signals | Battery degradation, Costly implementation | 1 – 10 sec | 15 min – 2 hrs |
| **Load Reduction** (Compensation NOT required in previous or subsequent hours) | TCL | HVAC | Global temperature adjustment | MILP, MINLP, SVR, ANN | EMCS, PCT, BAS | Ease of implementation, Potential energy efficiency improvement | Potential temperature discomfort of some users | 30 sec – 1 min | 1 – 2 hrs |
| | | Refrigeration | Temperature set point adjustment | MILP, MINLP, SVR, ANN | EMCS, PCT | Ease of implementation, Potential energy efficiency improvement | Risk of impact on product quality | 10 – 20 min | 30 min – 1 hr |
| | Industrial Processes | Oil Refinement | Mass flow regulation | MILP, SVR, ANN | EMCS, MPC | Potential energy efficiency improvements | Production limitation, Loss of potential revenue | 2 – 4 hrs | 2 – 4 hrs |
| | IT Industry | Data Centers | Power capping | RO, DP, RL | EMCS, SLURM | Capable of emergency responses, Ease of Implementation | Loss of productivity | 15 – 30 min | 2 – 4 hrs |
| **Load Removal** (100% Reduction) | Industrial Processes | Metalworking | Electric furnace temporary halt | MILP, ANN | EMCS | Fast response to DR signals, Large peak reduction potential | Short response duration, Complicated implementation, Risk of increased costs | 30 sec – 10 min | 30 min – 1 hr |

## 2.3. Energy Storage Systems

Energy storage systems (ESS) are practically built to serve as energy flexibility providers to the grid. They are capable of absorbing the intermittencies of different loads and generators, and thus, provide system stability [60]. In addition, ESS are also suitable for peak shaving services such as energy shifting, which help reduce renewable generation peaks or load demand peaks by absorbing excess power and supplying excess demand when necessary, thus reducing potential network congestion issues, as well as suboptimal contingency measures such as renewable energy curtailment [61], [62].

Some of the flexibility services that ESS can provide are based on the following:
- **Energy Shifting:** Scheduled charging and discharging operations to complement inflexible loads and/or generators by absorbing/releasing energy to primarily mitigate peaks.
- **Energy Regulation:** Quick response to instantaneous change in electrical demand and/or generation by quickly increasing or reducing charging/discharging rates, mitigating generation and load variability from inflexible resources.

The flexibility actions and all other characteristics of each ESS analyzed in this paper are presented based on how they can provide these services. Additionally, Table 4 summarizes all the information relevant to the flexible ESS discussed in this section that pertains to how they are operated to provide such services as energy regulation and shifting via generation and load intermittency mitigation and peak shaving, serving as devices that enhance flexibility of generation or load resources that are otherwise inflexible.

### 2.3.1. Electrochemical and Chemical Energy Storage

Electrochemical ESS types encompass all technologies that store their energy in chemical bonds that cause a potential difference, allowing direct extraction of energy in electrical form. Chemical energy storage, on the other hand, stores energy at a molecular level, in the form of a fuel, such as hydrogen. This fuel is later consumed to extract and convert the energy back into electrical form [63], [64].

#### 2.3.1.1. Battery Energy Storage

Due to their fast response and high energy density, battery energy storage systems (BESS) are one of the most applied energy storage technologies [65]. This technology is particularly effective at energy regulation, such as intermittency mitigation and load following actions, thanks to its fast response, but also very effective at providing energy shifting to suppress generation and load peaks by having a wide range of energy density, depending on the material composition of the battery [63]-[65]. These flexibility actions can be achieved by a BESS with the implementation of optimization methods such as GA, DP, and fuzzy logic control (FCL). These algorithms and optimization dynamics allow for an accurate synchronization with grid signals to actively respond to them in real time, depending on capacity, as well as to determine appropriate charge levels and charging and discharging schedules to supplement inflexible loads and generators [65]-[67]. These optimization strategies are also implemented thanks to supplemental technologies such as energy storage management software (ESMS) that allows programming and deployment of optimization models, AGC to regulate power output, and static synchronous compensators (STATCOM) to process the grid signals properly to maintain stability with their flexibility actions [67].

#### 2.3.1.2. Hydrogen Energy Storage

Hydrogen fuel cells convert hydrogen-rich fuel and air into electrical energy. Conversely, an electrolyzer produces hydrogen gas from water, which is later stored [68], [69]. Although not as efficient or as fast-responding as a BESS, hydrogen energy storage (HES) is still sufficiently fast to be used effectively for energy regulation services, in addition to boasting a much larger storage capacity due to its external storage in hydrogen storage tanks or salt caverns; thus allowing HES to provide energy shifting for more prolonged periods of time [63], [68].

Some of the optimization algorithms and strategies that make energy shifting possible for HES are optimization models such as MILP, bi-level MILP, and multi-objective particle swarm optimization (MOPSO) [69]. Moreover, for energy regulation such as intermittent generation and load demand changes, models such as FCL and machine learning methods like ANN can be used to better predict and actively follow sudden grid-changing conditions [70]. These actions are possible thanks to technologies such as MPC and EMCS modules that facilitate the implementation of these control strategies to achieve seamless synchronization with the grid and better provide flexibility [70].

*2.3.2. Mechanical Energy Storage*

Mechanical energy storage is one of the most versatile forms of energy storage. As such, it is a very popular and mature technology with the largest share of ESS in the world, with pumped hydro systems (PHS) accounting for around 67% of the total installed ESS capacity worldwide as of 2023; and even though BESS is rapidly growing and expected to surpass PHS, this technology is still expected to hold around 25% of the total global share by 2030 [71].

Mechanical energy storage can be in the form of potential energy, such as in the case of pumped hydro plants and compressed air, and in the form of kinetic energy, such as in the case of flywheels [63], [68].

*2.3.2.1. Flywheels*

Flywheels consist of a rotating cylindrical part and a magnetic suspension bearing assembly. The stored energy takes the form of kinetic energy. The flywheel's rotation is accelerated to transfer electrical energy to the stored kinetic energy, or decelerated to return the stored kinetic energy as electrical energy [68]. Flywheels have shown extremely high efficiencies for short-duration storage as well as very fast response times, which allows them to provide substantial flexibility when managing electrical disturbances in the grid. However, they suffer from high self-discharge rates [63], [64].

Energy regulation is offered by flywheels via variable generation intermittency mitigation, as well as load-following services for short-term responses. This is achieved with models such as the Markov decision process (MDP) and machine learning techniques like RL, which help implement responses that smooth the power output of variable RES such as wind power, as well as providing stability support [72]. These control methods are possible thanks to controllers like the active disturbance rejection controller (ADRC), which features quick feedback, accurate control, and robust resistance to disturbances [73].

*2.3.2.2. Pumped Hydro*

PHS is one of the most popular storage technologies due to its simplicity and large capacity in the range of 1 to 3000 MW. PHS is a mature and robust technology with a high efficiency of 76% – 85%, low capital cost per unit energy, a long storage period, practically unlimited life cycle, and a very long life of 50 years or more [63], [64]. They are good for both energy shifting and energy regulation services, such as peak shaving and load following actions, thanks to their fast-ramping capabilities and large-scale storage capacities [68].

There are many machine learning algorithms and optimization models that can be implemented to provide flexibility with PHS using MPC and EMCS technologies. Among these are DL, SCR, LSTM, and FLC, which help best determine when and how to respond to grid flexibility signals based on current parameters such as storage levels and required power output, and also predict appropriate generating/pumping schedules that would anticipate future grid statuses [74], [75].

*2.3.2.3. Compressed Air*

Compressed air energy storage (CAES) is the second-largest mechanical ESS technology for large-scale storage after PHS, thanks to its flexibility of capacity sizes, which makes it suitable for both long- and short-storage duration applications [64]. CAES plants consist of power train motors used to drive a compressor to compress air into a reservoir, a high- and low-pressure turbine, and a generator. They operate similarly to a conventional gas turbine, with the compression and expansion stages occurring independently or concurrently, depending on the plant type. During the compression stage, excess electricity is used to run a chain of compressors that inject air into the reservoir. Then, during the expansion stage, when electricity is required, the pressurized air is released from the reservoir and used to run a turbine to produce electricity [76].

AGC technologies are also used for CAES just like for gas turbine generators since their operation principle is the same, and these are used in conjunction with optimization models such as DP, and machine learning algorithms like ANN and DL to achieve speed and temperature control that allow for proper air compression and power generation to provide energy regulation as well as energy shifting services [77], [78].

*2.3.3. Thermal Energy Storage*

Thermal energy storage has been used for a long time for energy redistribution and energy efficiency improvement of thermal power plants on a short- and long-term basis [63], [64]. Thermal energy storage works by storing energy by heating or melting materials, with the energy now taking the form of sensible heat or latent heat [68].

*2.3.3.1. Latent Heat*

The heat absorbed or released by a body that only changes its physical state without changing its temperature is called latent heat. The amount of heat energy that can be stored by a body depends on its specific heat and heat capacity. The materials used in latent heat storage (LHS) are frequently referred to as PCMs. LHS systems are characterized by high energy density, and the energy storage capacity of the system is defined by the melting point and the enthalpy of state change of the PCM [63].

Table 4: Energy-flexible ESS [60]-[83].

| Flexibility Service | Resource Type | Flexibility Asset | Flexibility Action | Optimization Strategy | Supplement Technologies | Advantages | Disadvantages | Response Time | Response Duration |
|---|---|---|---|---|---|---|---|---|---|
| **Energy Shifting** (Requires compensation in previous or subsequent hours) | Electrochemical Energy Storage | BESS | Peak shaving, Excess generation absorption | FLC, DP | ESMS | Reduction of less efficient peaking plant use, Lower need for infrastructure upgrades | High capital costs, More challenging operation and maintenance | < 1 sec | < 12 hrs |
| | Chemical Energy Storage | HES | Peak shaving, Excess generation absorption | MILP, GA, MOPSO | MPC | Long-term energy storage capabilities, | Relatively low efficiency, Lack of robust hydrogen infrastructure | 1 – 10 sec | < 1 day |
| | Potential Energy Storage | PHS | Peak Shaving, Excess generation absorption | SVR, FLC | MPC, EMCS | Large storage capacity, long lifetime, Low maintenance costs | Environmentally constrained, Large unit size | 1 – 10 min | < 5 days |
| | | CAES | Peak Shaving, Excess generation absorption | ANN, BDL | AGC | Environmentally Friendly, Long lifetime | Geographically constrained | 1 sec – 1 min | < 1 day |
| | Thermal Energy Storage | LHS | Peak Shaving, Excess generation absorption | GA, MOPSO, LSTM | MPC, EMCS, PCM Encapsulation | High energy density, Small range of temperature and volume changes | High demand for PCMs, Slow response times | > 10 min | 1 - 24 hrs |
| | | SHS | Peak Shaving, Excess generation absorption | GA, ANN, PSO, MILP | MPC, EMCS, PCT | Low capital costs | Slow response times, low energy density | > 10 min | < 6 hrs |
| **Energy Regulation** (Quick adjustment in charging and discharging rates) | Electrochemical Energy Storage | BESS | Load following, Generation intermittency mitigation | FLC, GA | AGC, STATCOM | High reaction speed | Requires overcharge protection | < 1 sec | < 12 hrs |
| | Chemical Energy Storage | HES | Load following, Generation intermittency mitigation | FLC, ANN | ESMS, MPC | Scalability, Minimal environmental impact | Relatively low efficiency, Requires critical safety to handle hydrogen | 1 – 10 sec | < 1 day |
| | Mechanical Energy Storage | Flywheels | Load following, Generation intermittency mitigation | RL, MDP | EMCS, ADRC | High power density, low maintenance costs | Inefficient when idle | < 5 sec | 8 sec – 15 min |
| | | PHS | Load following, Generation intermittency mitigation | LSTM, SVR | MPC | High efficiency, Low maintenance costs, Fast Ramping capability | Large unit size | 1 – 10 min | < 5 days |
| | | CAES | Load following, Generation intermittency mitigation | DP, SVR, BDL | AGC | Fast reaction speed, Environmentally friendly | Low round-trip efficiency | 1 sec – 1 min | < 1 day |

The flexibility on energy storage capacity sizes allows LHS systems to be able to provide energy shifting services, such as generation and load peak shaving [79]. These services are possible thanks to optimization models that allow optimal scheduling under different conditions, like GA and MOPSO. Additionally, machine learning can also be used for forecasting grid conditions and better prediction of appropriate flexible actions, among such models can be ANN and DL techniques, which are implemented via MPC modules [80]-[82].

*2.3.3.2. Sensible Heat*

Sensible heat storage (SHS) is the simplest method of storing thermal energy. It stores energy by directly heating a solid or liquid medium without phase change. The energy storage density of SHS is mainly determined by the specific heat capacity of the storage material and the operating temperature range of the system [63], [68].

The concept is very similar to how LHS works, only with the main disadvantage that SHS presents a lower energy density, and that temperature variability is present as well [68]. Nonetheless, energy shifting can still be achieved by implementing GA or ANN algorithms for optimal demand matching and economic dispatch approaches, as well as PSO or MILP to handle optimal temperature levels that determine the amount of energy stored, and how to effectively store and release this energy [83].

## 3. Characterizing Energy Flexibility

There are four main parameters or characteristics that serve as good indicators of how much flexibility an energy resource can provide, or how well it can respond to grid-changing conditions to provide energy flexibility services. All these parameters pertain to how modifiable the resource's power output or input is [5], [11], [84]. These parameters have been identified as follows:
- **Power Operating Range:** The difference between the resource's maximum and minimum power operating levels.
  - For generators: maximum and minimum power output.
  - For loads: maximum and minimum power consumption levels.
  - For ESS: maximum and minimum charging/discharging rates.
  - Measured in units of power (i.e., W, kW, MW, etc.).
- **Power Ramping:** Rate at which power operating points can be increased or decreased over time.
  - For generators: power output ramp-up/ramp-down rates.
  - For loads: ramp-up/ramp-down of processes directly linked with power consumption
  - For ESS: ramp-up/ramp-down of charging/discharging rates.
  - Measured in units of power per time (i.e., W/s, kW/min, MW/min, MW/h, etc.).
- **Response Time:** Time it takes the resource to implement its flexibility action.
  - For generators: Time latency to begin output adjustment.
  - For loads: DR implementation time.
  - For ESS: Time to engage charging/discharging response.
  - Measured in units of time (i.e., sec, min, hours, etc.).
- **Response Duration:** Time duration of resource's response at its maximum operating point.
  - For generators: maximum operation time at maximum power output.
  - For loads: maximum time demand can be reduced to its lowest operating point.
  - For ESS: charging/discharging time at maximum power
  - Measured in units of time (i.e., sec, min, hours, etc.)

*3.1. Case Scenarios Illustrating Energy Flexibility Services*

When variable load and/or generation changes in an unforeseen manner, flexibility services must be implemented at either the supply or demand sides, or at both [3], [7]. Overall, any energy resources (i.e., generators, loads, ESS) could respond, if flexible enough, and provide any of the following services [12]:
- **Intermittency Mitigation**
  - Modifying power input/output to absorb intermittent variations in generation and/or load demand.
  - It mainly depends on the resource's power ramping capabilities and power operating range.
- **Peak Shaving**
  - Shifting load operations, generation, or charging/discharging at times that offset an invariable generation or load peak.
  - It mainly depends on power ramping capabilities, power operating range, and response duration.
- **Energy Reserve**

- o Operating at maximum/minimum base points so that a large amount of energy can be provided, or a large amount of load demand can be reduced in case of sudden and drastic changes in the supply-demand mix.
- o It depends on all four flexibility parameters: power ramping, power operating range, response duration, and response time.

For these services, different examples are set up to analyze how each resource is able to follow supply/demand variations under different scenarios. On each example for each resource, the amount of energy mismatch (or power deficit) between supply and demand is obtained and used to measure how flexible the resource's response is. Thus, the lower this mismatch between supply and demand, the more flexible the resource was in that scenario. These results are then used to find correlations between the resource's parameters and the amount of power deficit when meeting supply and demand needs.

*3.1.1. Intermittency Mitigation*

This example scenario simulates a case where there are sudden and unforeseen variations in supply and demand. These are in the form of intermittent increases and decreases of the "net load," which is the combined fixed generation and load that must be addressed (totally or partially) by the flexible resources. The scenarios for intermittency mitigation for loads, generators, and energy storage systems consist of two-hour periods that begin suddenly (i.e., without prior preparation), varying between positive and negative values with a maximum magnitude of 50% of the rated power of the resource being analyzed. The power magnitude of these net loads is normalized by the resource's rated power so that fair comparisons can be carried out among resources that have different power capacity sizes, allowing for scalability considerations.

The net load that is considered for all these cases of intermittency mitigation has the profile shown in Fig. 1.

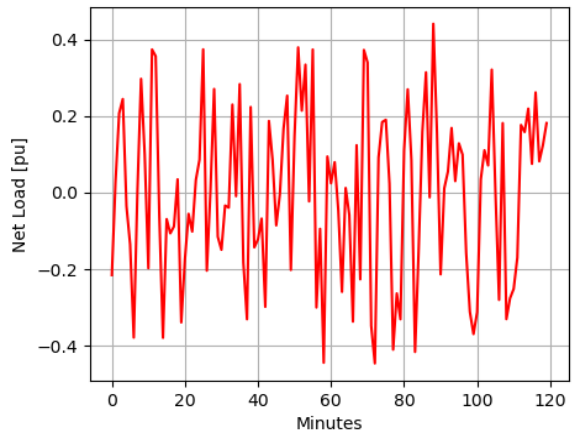

Fig. 1: Net load for intermittency mitigation scenarios.

The intermittency mitigation optimization model for each resource type is formulated as a time-independent optimization where the magnitude of the power deficit (i.e., the mismatch between power supply and demand) at every time interval is minimized individually. This is done to simulate a real-time response scenario where each time interval is not forecasted. The mathematical formulation of this real-time optimization model is included in Appendix A.

*3.1.2. Peak Shaving*

This scenario represents a case where there are large peaks of variable generation, such as those of solar power, and a large peak of load demand, which occurs at differing times, simulating what is more commonly known as the "duck curve" [85]. Thus, in this case, the resource is tested to see how well it can mitigate these peaks by shifting its operations or simply reducing or increasing its power output/input.

The example scenarios for peak shaving of loads, generators, and energy storage systems consist of 24-hour periods during which, in addition to the previously forecasted load and generation, there is also additional net load injection of peak invariable generation and demand. This particular net load profile is displayed in Fig. 2.

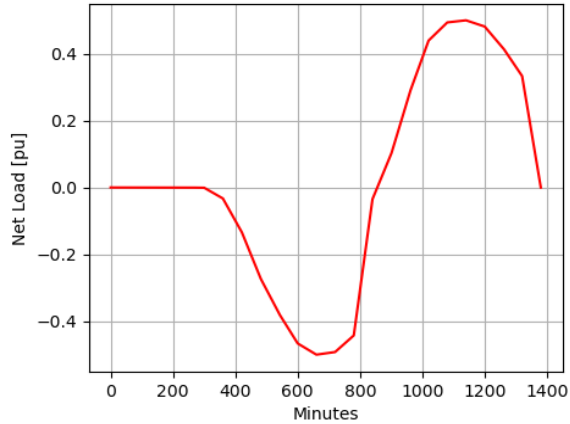

Fig. 2: Net load profile for peak shaving scenarios.

These peaks have a magnitude of half the rated power of the resource being analyzed, so that they can be used for comparison across all resource types when inspecting the amount of power deficit each exhibits when attempting to minimize these peaks.

The optimization model utilized for this scenario is essentially a day-ahead scheduling model where the optimization is time-dependent, which is meant to represent a scenario where large peaks are expected and only one resource can be rescheduled in an attempt to mitigate these peaks. The formulation of this optimization model is included in Appendix A.

### 3.1.3. Energy Reserve

This scenario represents a case where there are sudden, constant, and prolonged increases in load, as well as in generation. These increases are substantial in magnitude, and occur at separate times, testing how well a resource can respond to large net load deviations, both positive and negative. The scenarios for energy reserve for loads, generators, and energy storage systems will consist of 20-hour periods in which the first additional load is suddenly incorporated into the supply/demand mix, with a magnitude of half the rated power of the resource, and lasts for 6 hours before dropping back to zero. After another 6 hours then there is a sudden large increase in generation creating the same effect but now in a resulting "negative net load" for another 6 hours before going back to zero for the last two hours. This net load profile looks as shown in Fig. 3.

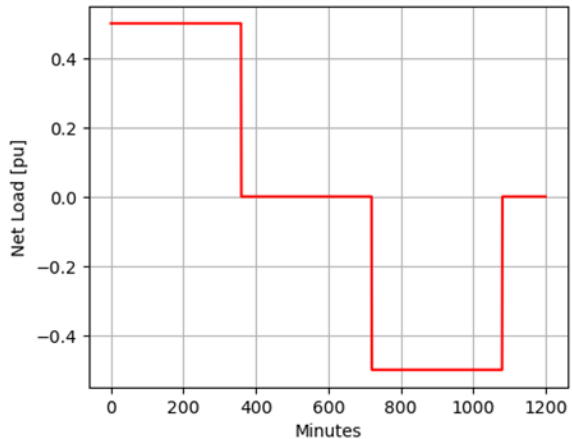

Fig. 3: Net load profile for peak shaving scenarios.

The optimization model to mitigate the power deficit of these cases is once again time-independent, to simulate the unforeseen nature of this example, where the resource has to respond and provide energy flexibility without prior knowledge and preparation. Therefore, the optimization model for these scenarios is the same as for the intermittency mitigation scenarios, and the detailed formulation is included in Appendix A.

### 3.2. Energy Flexibility of Generators

The generators in Table 5 are used to test and compare their energy flexibility in the three different scenarios previously defined. All scenarios assume that there has been a prior load that the generators were scheduled to serve, which falls within their capabilities, meaning that the load is fully met, and supply/demand is balanced before the unforeseen net load is injected.

Table 5: Parameters for different generator types [86]-[92].

| Gen. Type | Generator | Min. Power [MW] | Max. Power [MW] | Efficiency | Ramping Rate [MW/min] | Start-Up Time [min] |
|---|---|---|---|---|---|---|
| Thermal Generator | CCGT | 240 | 800 | 0.5 | 24 | 180 |
|  | ICE | 1.8 | 18 | 0.48 | 3.6 | 5 |
| Renewable Generator | Hydropower | 60 | 1900 | 0.9 | 50 | 1 |
|  | Solar PV System | 0 | 1.3 | 0.198 | 1000 | 1 |
|  | Wind Turbines | 0.0086 | 1.3 | 0.1626 | 2.6 | 1 |

The energy flexibility results for each generator type are shown in Tables 6, 7, and 8 for intermittency mitigation, peak shaving, and energy reserve, respectively. These results are in terms of power deficit (i.e., supply/demand mismatch) statistics, indicating how well each generator is at providing energy flexibility to minimize imbalances between supply and demand. The most representative quantity to measure flexibility is the root mean squared deficit, as this quantity captures not only the total energy mismatch of the entire period but also the individual magnitude at each time interval by squaring the absolute deficit.

Detailed plots showing the flexible performance of each generator are included as supplementary material and are described in Appendix B, which show generator power output compared against the overall net load and the power deficit results at every one-minute interval.

Table 6: Power deficit results statistics for generators under the intermittency mitigation scenario.

| Gen. Type | Generator | Average Absolute Deficit | Net Energy Deficit | Root Mean Squared Deficit |
|---|---|---|---|---|
| Thermal | CCGT | 0.1622 | 0.3245 | 0.2049 |
|  | ICE | 0.0712 | 0.1423 | 0.1222 |
| Renewable | Hydropower | 0.1645 | 0.3290 | 0.2067 |
|  | Solar Power | 0.1543 | 0.3085 | 0.2407 |
|  | Wind Power | 0.1228 | 0.2456 | 0.2106 |

Table 7: Power deficit results statistics for generators under the peak shaving scenario.

| Gen. Type | Generator | Average Absolute Deficit | Net Energy Deficit | Root Mean Squared Deficit |
|---|---|---|---|---|
| Thermal | CCGT | 0.0893 | 2.055 | 0.1369 |
|  | ICE | 0.0474 | 1.091 | 0.1010 |
| Renewable | Hydropower | 0.0426 | 0.9813 | 0.0993 |
|  | Solar Power | 0.1410 | 3.244 | 0.2368 |
|  | Wind Power | 0.1224 | 2.817 | 0.2156 |

Table 8: Power deficit results statistics for generators under the energy reserve scenario.

| Gen. Type | Generator | Average Absolute Deficit | Net Energy Deficit | Root Mean Squared Deficit |
|---|---|---|---|---|
| Thermal | CCGT | 0.0564 | 1.270 | 0.0973 |
|  | ICE | 0.0319 | 0.6378 | 0.0611 |
| Renewable | Hydropower | 0.0443 | 0.8866 | 0.0908 |
|  | Solar Power | 0.2706 | 5.411 | 0.3643 |
|  | Wind Power | 0.2059 | 4.119 | 0.2834 |

As observed, the generator that provides the most flexibility when it comes to addressing sudden variations in supply and demand is the internal combustion engine. This generator performed well at mitigating quick intermittent variations, as well as large and prolonged variations, and energy shifting, achieving generation and demand peak shaving. Moreover, only the hydropower generator performed comparably to the ICE in peak shaving, providing a slight improvement in the net energy mismatch. Therefore, an ICE generator seems to be a much more suitable generator to provide energy flexibility in all scenario types. Moreover, it is important to mention that this is achieved by ICE generators operating with liquid fuels such as diesel to achieve the listed ramping rate, as gas-based fuels may restrict what the effective ramping of the unit may be.

*3.3. Energy Flexibility of Loads*

The loads in Table 9 are used to test and compare their energy flexibility in the three different scenarios previously stated. All scenarios assume that there has been a prior generation scheduled to serve a previously forecasted demand for these loads under normal operation, thus having supply and demand fully balanced before the unforeseen net load is injected.

The energy flexibility results for each load type are shown in Tables 10, 11, and 12 for intermittency mitigation, peak shaving, and energy reserve, respectively. The root mean squared deficit is again used to estimate flexibility in terms of overall energy deficit as well as power deficit minimization at every time interval.

Moreover, detailed plots showing the flexible performance of all these loads are also included as supplementary material and are described in Appendix B, which show load power consumption compared against the overall invariable generation available, and the power deficit result at every one-minute interval.

Table 9: Parameters for different types of loads [36], [40], [48], [49], [55], [59], [93]-[99].

| Load Type | Load | Min. Power Consumption [kW] | Max. Power Consumption [kW] | Energy Intensity | Power Ramping [kW/min] | Response Time [min] |
|---|---|---|---|---|---|---|
| TCL | Refrigeration | 180 | 360 | 31596.66 [kWh/°C] | 180 | 10 |
|  | HVAC | 4.5 | 7.2 | 2.5 [kWh/°C] | 7.2 | 1 |
| Industrial Process | Cement Production | 138 | 2370 | 3.58 [kWh/ton] | 27.18 | 10 |
|  | Oil Refinement | 25,000 | 35,000 | 0.025 [kWh/kg] | 83.33 | 240 |
| IT Industry | Data Center | 1250 | 5000 | $5.29 \times 10^{-13}$ [Wh/CPU cycle] | 333.33 | 15 |

Table 10: Power deficit results statistics for loads under the intermittency mitigation scenario.

| Load Type | Load | Average Absolute Deficit | Net Energy Deficit | Root Mean Squared Deficit |
|---|---|---|---|---|
| TCL | Refrigerated Warehouse | 0.2244 | 0.4487 | 0.3190 |
|  | HVAC | 0.1699 | 0.3398 | 0.2009 |
| Industrial Process | Cement Plant | 0.1740 | 0.3481 | 0.2151 |
|  | Oil Refinement Plant | 0.1929 | 0.3858 | 0.2457 |
| IT Industry | Data Center | 0.1460 | 0.2919 | 0.1979 |

Table 11: Power deficit results statistics for loads under the peak shaving scenario.

| Load Type | Load | Average Absolute Deficit | Net Energy Deficit | Root Mean Squared Deficit |
|---|---|---|---|---|
| TCL | Refrigerated Warehouse | 0.0595 | 1.368 | 0.1338 |
|  | HVAC | 0.2335 | 5.374 | 0.3052 |
| Industrial Process | Cement Plant | 0.2206 | 5.077 | 0.3099 |
|  | Oil Refinement Plant | 0.4263 | 9.812 | 0.5152 |
| IT Industry | Data Center | 0.2103 | 4.841 | 0.3047 |

Table 12: Power deficit results statistics for loads under the energy reserve scenario.

| Load Type | Load | Average Absolute Deficit | Net Energy Deficit | Root Mean Squared Deficit |
|---|---|---|---|---|
| TCL | Refrigerated Warehouse | 0.0354 | 0.7083 | 0.1331 |
|  | HVAC | 0.1899 | 3.798 | 0.2828 |
| Industrial Process | Cement Plant | 0.0514 | 1.028 | 0.1376 |
|  | Oil Refinement Plant | 0.3605 | 7.210 | 0.4686 |
| IT Industry | Data Center | 0.0048 | 0.0953 | 0.0392 |

In the case of these loads, it can be observed that the refrigerated warehouse showed the most flexibility when reducing peaks by redistributing its consumption more uniformly over the time period. However, when evaluating flexibility to mitigate sudden and sporadic fluctuations in the supply-demand mix, as well as sudden but prolonged increases in supply and demand, the data centers were the type of load to outperform the others by a substantial amount of power deficit reduction. This means that data centers are a more versatile load that can modify their output quickly and substantially, but refrigerated warehouses are better at rescheduling and redistributing their consumption more uniformly while still achieving very similar performance. Moreover, these results also reveal that the oil refinement load type is the least flexible load of all three scenario types. Therefore, this type of load would be one that may be considered relatively inflexible and unsuitable for DR or DSM services.

*3.4. Energy Flexibility of Energy Storage*

The ESS shown in Table 13 are used to test and compare their energy flexibility in the three different scenarios previously stated. All scenarios assume both generation and load have been priorly scheduled to balance each other out, leaving the new net loads that are injected as the only variable load and generation that is introduced to be met solely by the actions of the energy storage devices.

The energy flexibility results for each ESS type are shown in Tables 14, 15, and 16 for intermittency mitigation, peak shaving, and energy reserve, respectively. These results are also obtained in terms of power deficit statistics, however, now with the peculiar ability of ESS to mitigate variations in both power supply and power demand in the grid. These results also serve to indicate how well each ESS can handle energy flexibility requests to minimize imbalances between supply and demand through their charging and discharging operations.

Detailed plots showing the flexible performance of all these ESS are also included as supplementary material and are described in Appendix B, which show ESS charging and discharging rates compared against the overall net load they are meant to mitigate, and the power deficit result at every one-minute interval.

Table 13: Parameters for three different types of ESS [61], [63], [64], [68], [72], [73], [88], [100]-[106].

| ESS Type | ESS | Min. Power [MW] | Max. Power [MW] | Energy Capacity [MWh] | Charging Efficiency | Discharging Efficiency | Power Ramping [MW/min] | Start-Up Time [min] |
|---|---|---|---|---|---|---|---|---|
| Electrochemical Energy Storage | Battery | 0.1 | 100 | 400 | 0.9 | 0.97 | 6000 | 0 |
| Mechanical Energy Storage | Pumped Hydro | 100 | 5,000 | 8,000 | 0.7 | 0.85 | 50 | 1 |
|  | Flywheel | 0 | 1.0 | 0.25 | 0.98 | 0.98 | 15 | 0 |
| Thermal Energy Storage | Latent Heat | 0.1 | 300 | 2,500 | 0.75 | 0.90 | 0.48 | 60 |

Table 14: Power deficit results statistics for ESS under the intermittency mitigation scenario.

| ESS Type | ESS | Average Absolute Deficit | Net Energy Deficit | Root Mean Squared Deficit |
|---|---|---|---|---|
| Chemical | Battery | 0.0839 | 0.1678 | 0.1439 |
| Mechanical | Pumped Hydro | 0.1718 | 0.3435 | 0.2128 |
|  | Flywheel | 0.0839 | 0.1678 | 0.1439 |
| Thermal | Latent Heat | 0.1837 | 0.3674 | 0.2212 |

Table 15: Power deficit results statistics for ESS under the peak shaving scenario.

| ESS Type | ESS | Average Absolute Deficit | Net Energy Deficit | Root Mean Squared Deficit |
|---|---|---|---|---|
| Chemical | Battery | 0.0282 | 0.6488 | 0.0464 |
| Mechanical | Pumped Hydro | 0.0924 | 2.126 | 0.1274 |
|  | Flywheel | 0.2291 | 5.272 | 0.2887 |
| Thermal | Latent Heat | 0.1209 | 2.782 | 0.1651 |

Table 16: Power deficit results statistics for ESS under the energy reserve scenario.

| ESS Type | ESS | Average Absolute Deficit | Net Energy Deficit | Root Mean Squared Deficit |
|---|---|---|---|---|
| Chemical | Battery | 0.0001 | 0.0167 | 0.0204 |
| Mechanical | Pumped Hydro | 0.1177 | 2.354 | 0.2352 |
|  | Flywheel | 0.2750 | 5.500 | 0.3707 |
| Thermal | Latent Heat | 0.2514 | 5.028 | 0.3006 |

Comparing these three different types of ESS, it can be observed that for the mitigation of quick and sporadic fluctuations in the supply-demand mix, both flywheels and batteries are the most suitable technologies. However, for energy shifting services as well as providing prolonged and continuous energy reserve services, BESS are far superior to any other energy storage technology due to their rapid ramping capabilities and 4-to-1 energy-to-power ratio. It is important to note that while BESS have been identified in the literature to possess lower energy to power ratios in terms of their energy and power densities [61], [68], [100], [101], configurations with either a 2-to-1 or a 4-to-1 energy-to-power ratio are most common [107], as is the case with Tesla's Megapack and Sungrow PowerTitan BESS modules [102], [108]. This is largely due to inverter configuration and power rating limitations, as well as potential regulations and policies on capacity and storage requirements.

The second most flexible ESS technology for peak shaving and reserve applications is pumped hydro plants, which are more suitable in applications that require large amounts of energy storage and power capacity. Latent heat is also a good candidate for peak shaving, although pumped hydro still offers more flexibility for that type of service.

## 4. Conclusions

Energy flexibility is the ability of different energy resources to adjust their electrical supply and/or demand in a timely manner to maintain a balance between supply and demand. Energy flexibility services can be provided by any type of energy resource connected to the electrical grid, so long as they can be controlled to a certain extent. Generators provide energy flexibility by adjusting their electrical power output in response to changing demand needs in the grid, loads can provide flexibility by adjusting their power consumption according to changing conditions, and energy storage systems act as a flexibility buffer for both generators and loads so that additional supply or demand can be incorporated to complement the less controllable generators and loads.

This paper reviewed a number of different generators, loads and energy storage types, categorized them, and explored their modes of operations and capabilities to provide flexibility under different circumstances, as well as the control strategies and technologies that enable them to respond to grid-changing conditions, how fast they can provide these services and for how long. Within the generator category, both thermal and renewable generators were considered, with two variable renewable energy sources explored, which, however limited, can still offer a certain amount of flexibility if operated in special ways, such as the case of wind and solar power performing curtailment. In the load category, different loads such as TCLs, industrial processes, and

IT industry loads like data centers were considered; and for ESS, thermal, chemical, and mechanical energy storage resources were all considered and compared.

While the list of resources considered in this paper is not exhaustive, it incorporates a variety of different types for each category in order to capture the most common features within different resources that allow them to operate flexibly. In addition, some of the resources considered and compared were also utilized in three different scenarios to measure how much flexibility they would be able to provide under different circumstances based on their operating range, ramping rate, response time, and response duration. The three different scenarios were (i) intermittency mitigation, which simulates a scenario in which the supply and demand mix suddenly and sporadically increases and decreases randomly, (ii) peak shaving, which simulates scenarios with peak uncontrollable load and generation at different times of a 24-hour period, and (iii) energy reserve, which simulates a scenario where the supply-demand mix dramatically increases suddenly, followed by a sudden drastic decrease as well. These scenarios show which resources are more appropriate for each of them and provide a frame of reference to construct more comprehensive comparisons considering more types of generators, loads, and ESS.

**Acknowledgments**

This work was (in part) funded by Shell. It is based upon collaborative work between the University of Houston and Shell. The authors would like to acknowledge Elizabeth Endler, Chief Scientist - Power and System Integration, and Fred Drewitt, Head of System Integration & Flex Tech, at Shell International Exploration & Production Inc, for their valuable insight, support, and expertise comments when preparing this paper.

**Appendix A. Optimization Models**

This appendix includes the mathematical formulation of the optimization models that are used in Section 3 and that are used to derive the results of each corresponding scenario.

Table A. 1: Nomenclature for the optimization models.

| Symbol | Definition |
|---|---|
| $t$ | Time interval index. |
| $T$ | Set of time intervals (i.e., time period) |
| $\Delta t$ | Length of each time interval |
| $u_t$ | Resource status indicator at every time interval. |
| $OFF_t$ | Total time of resource being offline at every time interval. |
| $t^{SU}$ | Resource's start-up time. |
| $P_{max}$ | Resource's maximum power input/output. |
| $P_{min}$ | Resource's minimum power input/output. |
| $r$ | Resource's ramping rate. |

*A.1. Intermittency Mitigation and Energy Reserve Model*

This optimization model involves constraints that place limitations on the power output/input of the resources according to their parameters, such as minimum and maximum power, power ramping, response time, etc. As observed in the model, the optimization is performed for every time interval independently, emulating a real-time response.

Time-Independent Optimization

$$\begin{aligned}
&\text{minimize} \quad \left(P_t^{offset}\right)^2 \\
&\text{subject to} \quad \textbf{If } t \geq t^{SU} \\
&\qquad \textbf{If } \sum_{\tau=t-t^{SU}}^{t-1}(u_\tau) = 0 \text{ OR } u_{t-1} = 1 \\
&\qquad\qquad P_{min}\, u_t \leq P_t \leq P_{max} u_t \\
&\qquad \textbf{Else} \\
&\qquad\qquad P_t = 0 \\
&\qquad\qquad u_t = 0 \\
&\quad \textbf{Else} \\
&\qquad \textbf{If } u_0 = 0 \\
&\qquad\qquad P_t = 0\,, \forall t \in T \\
&\qquad\qquad u_t = 0\,, \forall t \in T \\
&\qquad \textbf{Else} \\
&\qquad\qquad \textbf{If } t > 0 \text{ AND } u_{t-1} = 0 \\
&\qquad\qquad\qquad P_t = 0\,, \forall t \in T \\
&\qquad\qquad\qquad u_t = 0\,, \forall t \in T \\
&\qquad\qquad \textbf{Else} \\
&\qquad\qquad\qquad P_{min} u_t \leq P_t \leq P_{max} u_t
\end{aligned}$$

$$u_t \in \{0,1\}$$
$$P_t - P_{t-1} \leq \Delta t \cdot r \cdot u_{t-1} + P_{min}(u_t - u_{t-1})$$
$$P_{t-1} - P_t \leq \Delta t \cdot r \cdot u_t + P_{min}(u_{t-1} - u_t)$$

This model is used for all resource types (generators, loads, ESS), and the only component that changes is the power balance constraint. This is because the role of the power variable changes from load to generation, depending on the resource and its operating mode. For generators, the power balance constraint will be $P_t = P_t^L + P_t^{net} - P_t^{offset}$, where $P_t$ is the generator power output, $P_t^L$ is the previously scheduled and forecasted load for the following time intervals, $P_t^{net}$ is the intermittent net load introduced, and $P_t^{offset}$ is the amount of power that could not be provided by the generator to fully meet the new load. For loads, the power balance constraint will be $P_t^{GEN} = P_t + P_t^{net} - P_t^{offset}$, where $P_t$ now represents the load's power consumption, and $P_t^{GEN}$ is the power generation priorly scheduled to meet the forecasted demand of the load, not accounting for the sudden intermittent net load represented by $P_t^{net}$. For ESS, the power balance constraint will be $P_t = P_t^{net} - P_t^{offset}$, where $P_t$ is now the power of the energy storage unit, which is positive when discharging, and negative when charging. In this case, there are no prior scheduled demand or generation factored into the constraint because it is assumed these were already scheduled as fixed demand and generation balancing each other out, leaving the energy storage unit to fully mitigate the intermittent net load on its own.

*A.2. Peak Shaving Model*

This is a time-dependent optimization model that essentially performs energy shifting to reduce large, invariable generation and load peaks by redistributing controllable generation and load.

Time-Dependent Optimization

$$\text{minimize} \quad \sum_{t \in T}(P_t^{offset})^2$$
$$\text{subject to}$$
$$P_{min} u_t \leq P_t \leq P_{max} u_t, \forall t \in T$$
$$OFF_t = (1 - u_{t-1})(OFF_{t-1} + 1), \forall t \in T$$
$$u_t - u_{t-1} \leq \frac{OFF_t}{t^{SU}}, \forall t \in T$$
$$u_t \in \{0,1\}, \forall t \in T$$
$$P_t - P_{t-1} \leq \Delta t \cdot r \cdot u_{t-1} + P_{min}(u_t - u_{t-1}), \forall t \in T$$
$$P_{t-1} - P_t \leq \Delta t \cdot r \cdot u_t + P_{min}(u_{t-1} - u_t), \forall t \in T$$
$$P_t^{GEN} = P_t^L + P_t^{net} - P_t^{offset}, \forall t \in T$$

The power balance constraints take the same form as those for the intermittency mitigation scenarios depending on the resource type (generators, loads, energy storage), only that now the constraint is for all time intervals in the 24-hour period rather than for a single time interval individually.

Moreover, the following constraints are added for the peak shaving scenarios to restrict the power deficit exclusively to when the peaks take place, depending on the resource type.

Power Balance for Generators

**If** $P_t^L + P_t^{net} < 0$
$$P_t^L + P_t^{net} \leq P_t^{offset} \leq 0, \forall t \in T$$
**Else**
$$0 \leq P_t^{offset} \leq P_t^L + P_t^{net}, \forall t \in T$$

Power Balance for Loads

**If** $P_t^{GEN} - P_t^{net} > 0$
$$-(P_t^{GEN} - P_t^{net}) \leq P_t^{offset} \leq 0, \forall t \in T$$
**Else**
$$0 \leq P_t^{offset} \leq -(P_t^{GEN} - P_t^{net}), \forall t \in T$$

Power Balance for ESS

**If** $P_t^{net} < 0$
$$P_t^{net} \leq P_t^{offset} \leq 0, \forall t \in T$$
**Else**
$$0 \leq P_t^{offset} \leq P_t^{net}, \forall t \in T$$

These constraints are added because one of the objectives of peak shaving is to reduce the magnitude of the peaks rather than only reduce the overall energy deficit over the 24-hour period, where these invariable peaks are introduced.